# PERSPECTIVES OF SUPERCONDUCTING MgB$_2$ FOR MICROWAVE APPLICATIONS

Matthias A. Hein

*Dept. of Physics, University of Wuppertal, Gauss-Strasse 20, D-42097 Wuppertal, Germany.*
*E-mail: mhein@venus.physik.uni-wuppertal.de*

**ABSTRACT**

We discuss the temperature, frequency, and power-dependent surface resistance of the boride superconductor MgB$_2$ in relation to possible applications for passive microwave devices. The data available in the literature are compared with results for polycrystalline Nb$_3$Sn and epitaxial YBa$_2$Cu$_3$O$_{7-x}$, which are representative of the classical and cuprate superconductors. MgB$_2$ displays all specific features that make superconductors attractive for high-performance devices, even though the fabrication technology is not yet mature. We attempt to identify promising areas of applications, as well as material requirements, which could further promote the attractiveness of the new superconductor in this field.

**INTRODUCTION**

The recently discovered 40K superconductor MgB$_2$ [1] (review in [2]) has received great attention because of the simple binary chemical composition, the high transition temperature $T_c$ compared to other metallic superconductors like Nb$_3$Sn, and the apparent BCS-like superconducting behaviour. The latter implies an exponentially vanishing density of normal electrons at low temperatures and correspondingly low power dissipation. The key advantages of MgB$_2$ over the high $T_c$ superconductors like YBa$_2$Cu$_3$O$_{7-x}$ (YBCO) are the the longer coherence length, which makes the material less susceptible to structural defects like grain boundaries, and the smaller anisotropy, which opens the potential for three-dimensional devices. It has been shown that good films can be deposited on a variety of substrates, including metals, without the need for epitaxial growth [2-4]. While the detailed theoretical understanding of superconductivity in MgB$_2$ is not yet complete and the material is still inhomogeneous (see below), the available data already provide sufficient ground to consider perspectives for applications in superconducting electronics. Such applications concern, e.g., Josephson devices and passive microwave components, both of which could benefit from the high gap frequency, $2\Delta/h$ around 1-2 THz [5,6], in a temperature range that is readily accessible with closed-cycle cryocoolers.

**POTENTIAL OF SUPERCONDUCTORS FOR MICROWAVE DEVICES**

Superconductors enable the realisation of high-performance RF and microwave devices [7-10], due to the low power dissipation, frequency independent penetration depth and the steep transition between the superconducting and the normal state. Related benefits are high energies stored per period, and hence high unloaded quality factors of resonators, strong miniaturisation, and multi-function integration, as well as wideband operation and high sensitivity. Examples include ultra sensitive detectors for radio astronomy, high field-gradient particle accelerators, miniaturised filter banks with extremely sharp bandpass characteristics, dispersive delay lines with high bandwidth-delay-products, and sensitive RF coils for magnetic resonance receivers for biomedical or pharmaceutical applications (MRI and NMR).

These advantages come, however, at the expense of a cryogenic environment and the related increase in fabrication or maintenance costs compared to conventional technology. The competitiveness of superconductors for microwave applications relies, therefore, on cheap substrates and preparation techniques and the highest possible operating temperatures. In this context, MgB$_2$ offers a challenging alternative to the low and high temperature superconductors (LTS, HTS).

**MICROWAVE MEASUREMENTS OF SUPERCONDUCTORS**

The microwave response of superconductors can be studied through measurements of the temperature ($T$), frequency ($f$), and microwave field ($B_s$) dependent surface impedance $Z_s = R_s + jX_s$ [8]. $Z_s$ represents the wave impedance of the material, $Z_s = \sqrt{\mu/\varepsilon}$, with $\mu$ the magnetic permeability and $\varepsilon = \sigma/j\omega$ the effective permittivity. The dynamic conductivity $\sigma$ provides information on the phase purity and electronic perfection of the material. Typical experimental techniques are based on measuring the quality factor and changes of the centre frequency of resonators, which are partly or entirely composed of the material under test. Knowing the distribution of the electric and magnetic fields in the resonator, from numerical simulations or reference experiments, $Z_s(T,f,B_s)$ can then be derived from the measured quantities. The electrodynamic response above 100 GHz is usually studied by quasi-optical techniques and infrared spectroscopy (e.g., [6]).

**MICROWAVE PROPERTIES OF MgB$_2$**

Numerous groups have studied the surface impedance of MgB$_2$ wires, pellets and polycrystalline, c-axis oriented, films, mainly on sapphire and LaAlO$_3$ substrates (e.g., [11,12] and references therein). We attempt to summarise the major results of these activities, including our own measurements at 87 GHz using MgB$_2$ films with high transition temperatures ~37.5 K. These films were prepared at LG Elite by Mg-vapour diffusion into a previously evaporated B-layer [3].

**Temperature Dependence of Surface Resistance and Penetration Depth**

Fig. 1 compares typical $R_s(T)$ data of a 400 nm thick polycrystalline MgB$_2$ film, before and after ion milling, with results for polycrystalline Nb$_3$Sn on sapphire ($T_c$=18 K) and epitaxial YBCO on MgO ($T_c$=90 K), on a reduced temperature scale, $t=T/T_c$. The transition into the superconducting state is much more gradual for MgB$_2$ than for the LTS and HTS films. Such a gradual transition is expected if impurity phases with a range of $T_c$-values are present, or if the superconductor displays a small energy gap. Both mechanisms could be relevant for MgB$_2$ (see below). According to the normal skin effect, the $R_s(T_c)$-value corresponds to a resistivity $\rho$~20 µΩcm, which is typical for high-quality films, but higher by about a factor of 10 than for single-crystals. The corresponding values for Nb$_3$Sn and YBCO films are ~8 and ~100 µΩcm. The residual resistance of the MgB$_2$ film of about 3.5 mΩ reflects, most likely, the presence of normal-conducting or lossy insulating impurities (e.g., Mg or MgO [3,13]), which could also be responsible for the slight upturn of $R_s$ at low temperature [14]. Upon thinning the film by about 50 nm, the surface resistance decreased in the superconducting state, indicating the removal of a lossy surface layer. The absolute level of the surface resistance even approached that of the best epitaxial YBCO films at a corresponding value of the reduced temperature. We observed a $T_c$ degradation of about 2.5 K, and an increase of the normal-state resistivity to about 30 µΩcm. The ion bombardment may have introduced additional disorder, thus reducing the electronic mean free path and suppressing the superconducting order parameter locally. Thus there may be remaining potential to reduce the $R_s$-level further.

Another feature seen in Fig. 1 is the variation of $R_s(T)$ at intermediate temperatures, $t$~0.4-0.8. While the surface resistance of Nb$_3$Sn can be well described by a large energy gap $\Delta/kT_c$~2.1 [15], the broad plateau of $R_s(T)$ for the YBCO film reflects the strong quasiparticle scattering as well as the d-wave symmetry of the order parameter in the cuprates [16]. The temperature dependence of $R_s$ of MgB$_2$, as well as that of changes of the penetration depth $\lambda$, is consistent with a small energy gap, $\Delta/kT_c$~1 (e.g., [11,12,17] and references therein). It is not yet clear why high-frequency measurements seem not to sense the larger gap value, $\Delta/kT_c$~2, which was observed in ac measurements of $\Delta\lambda(T)$ [18,19], tunneling and specific heat data [2]. Possible explanations could be related to tunneling (or proximity coupling) across insulating (or metallic) barriers between adjacent, slightly misaligned, grains (e.g., [5]). The steepness of $R_s(T)$, and thus the potential of MgB$_2$ for high-frequency applications, should then improve upon optimising phase purity and texture.

**Frequency Dependence of the Surface Resistance**

The suitability of MgB$_2$ for microwave applications depends, beside the temperature variation of $R_s$ and $\lambda$, on their frequency dependences. One expects $R_s(f) \propto f^2$ and $\lambda(f) \propto f^0$, unless normal electrons contribute to the shielding of the electromagnetic field [8]. This happens if $f$ approaches $2\Delta/h$ or the electronic relaxation rate, $\Gamma \sim \rho/\mu_0\lambda^2$, or if non-super-conducting phases are present. Fig. 2 compares $R_s(f)$-data for MgB$_2$, LTS, HTS, and cryogenic copper. The shaded regions indicate ranges of temperatures. The data for Nb$_3$Sn, and for YBCO at 77 K, follow the expected $f^2$-behaviour. The broad band of $R_s$-values covered by Nb$_3$Sn between 4.2 and 9 K reflects the exponential $R_s(T)$, while the narrower band for YBCO indicates the flatter $R_s(T)$-dependence and the higher $R_s(T\to 0)$-values (Fig. 1). The $R_s(f)$-data for MgB$_2$ also follow a quadratic dependence, at $T=4.2$ K on a similar $R_s$-level as for epitaxial YBCO, but only for the best samples [11,12]. Quasi-optical measurements indicate a gap frequency around 730 GHz [6] (arrow in Fig. 2). The surface resistance of the best MgB$_2$ samples remains below that of Cu up to ~400 GHz at 4.2 K and ~100 GHz at 30 K, while polycrystalline bulk material may exceed this limit already at about a factor of 5 lower frequencies. Further improvements in sample quality would extend the accessible range of frequencies and operating temperatures.

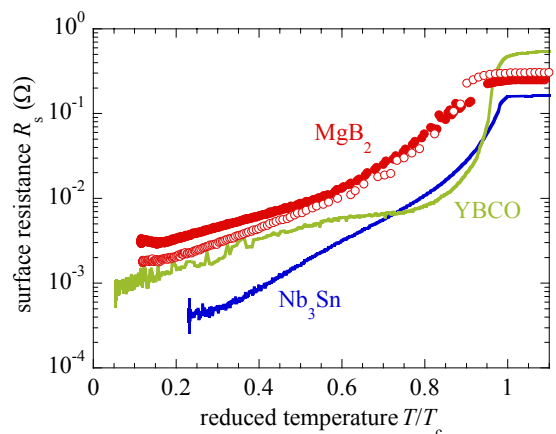

Fig. 1. Temperature dependence of $R_s$ at 87 GHz for a MgB$_2$ film before (solid) and after ion-beam milling (open red symbols), and for polycrystalline Nb$_3$Sn (blue, [15]) and epitaxial YBCO (green, [16]). All data were corrected for finite film thickness.

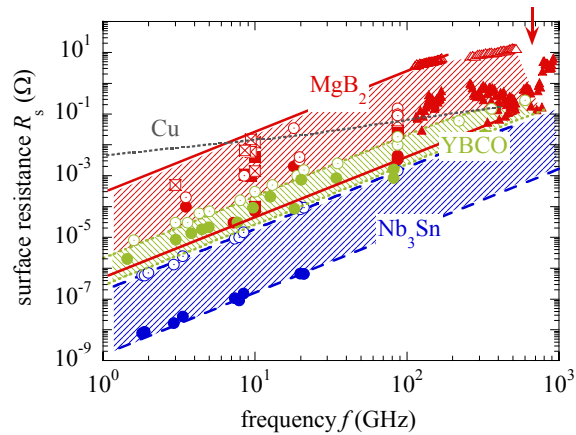

Fig. 2. Frequency dependence of $R_s$ for MgB$_2$ bulk (red squares) and films (red circles) at 4.2 K (solid) and 30 K (open symbols) [6,12,20], as well as for polycrystalline Nb$_3$Sn at 4.2 and 9 K (blue), epitaxial YBCO at 4.2 and 77 K (green), and copper at 77 K (grey) [8]. The arrow indicates the gap frequency of MgB$_2$, $2\Delta/h \sim 730$ GHz [6].

**Nonlinear Microwave Response**

Some microwave components, like particle accelerators or narrow-band filters, involve high oscillating powers [7,8]. It is important in such cases that the surface resistance (and reactance) remain independent of the microwave field amplitude. For superconductors, the linear microwave response can be limited by various mechanisms. A microscopic limit is given by the lower critical field $B_{c1}$, above which magnetic flux may enter the superconductor and lead to enhanced dissipation. Typical values for MgB$_2$ are $B_{c1}(T=0)\sim 25-50$ mT [2] (corresponding to a surface current $H_s\sim 20-50$ kA/m and a critical current density $J_c=H_s/\lambda\sim 15$ MA/cm$^2$), ~140 mT for Nb$_3$Sn and ~80 mT for YBCO [8]. These values decrease with increasing temperature, roughly like $1-t^2$. Additional limitations can be caused by heating and weak grain boundary coupling. Fig. 3 displays $R_s(B_s)/R_s(0)$ data for the three superconductors under various conditions. While the LTS and HTS have been studied for a long time [8,10], only few data are available for MgB$_2$ at present [11,21]. The surface resistance of Nb$_3$Sn and YBCO becomes field dependent at $B_s$=10-50 mT at low temperatures, $t$<0.3. $B_s$~$B_{c1}$ has been approached only for films with optimized grain structure and stoichiometry, and under optimized cooling conditions [8, 10,14,15]. The first $R_s(B_s)$ data for MgB$_2$ are still worse than for HTS and LTS, but promising in view of the moderate quality of the studied films ($T_c$= 26 and 33.5 K [11,21]), and given that field levels of ~10 mT will be sufficient for most applications. MgB$_2$ may become of interest even for particle accelerators, if high-quality films can be deposited onto high thermal-conductivity metallic substrates. The potential benefit would be a higher operating temperature compared to Nb$_3$Sn or even Nb, despite the low $B_{c1}$-value, which would limit the achievable field-gradients to about 10 MV/m.

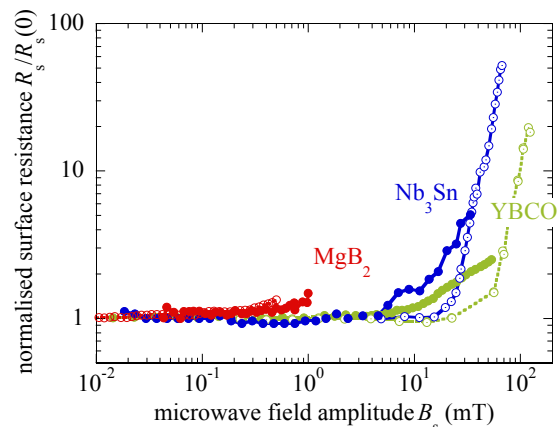

Fig. 3. Microwave field dependence of $R_s$ for MgB$_2$ on sapphire (red/solid: 10 GHz, $t=0.6$, $R_s(0)=1$ mΩ, [11]; red/open: 20 GHz, $t=0.15$, $R_s(0)=0.4$ mΩ, [21]), for a polycrystalline Nb$_3$Sn film on sapphire (blue/solid: 19 GHz, $t=0.2$, $R_s(0)<25$ μΩ, [15]), an accelerating cavity with Nb$_3$Sn on Nb (blue/open: 1.5 GHz, $t=0.2$, $R_s(0)\sim 12$ nΩ, [22]), and epitaxial YBCO films on sapphire (green/solid: unpatterned, 19 GHz, $t=0.05$, $R_s(0)\sim 110$ μΩ, [23]) and MgO (green/open: patterned, 2.3 GHz, $t=0.2$, $R_s(0)\sim 2.5$ μΩ, [24]).

**SUMMARY**

The temperature, frequency, and power dependent surface impedance of superconducting MgB$_2$ has been analysed. The surface resistance of polycrystalline MgB$_2$ films is comparable with that of epitaxial YBCO at similar reduced temperatures. $R_s$ stays below that of Cu up to mm-wave frequencies in a temperature range accessible with closed-cycle cryocoolers. The nonlinear response is limited, at present most likely due to impurities, to field levels well below $B_{c1}$. However, the achieved performance is promising for RF and microwave device applications. An obvious advantage over LTS is the higher $T_c$. The larger coherence length and lower anisotropy make MgB$_2$ competitive to HTS, especially if 3-dimensional devices like antennas, or low-cost fabrication, are of concern. Further improved phase purity, texture, and chemical stability, and the possibility to use cheap dielectric or metallic substrates are desirable, and also seem possible.

**ACKNOWLEDGMENTS**

The MgB$_2$ films measured at Wuppertal were provided by H.N. Lee and S.H. Moon. We acknowledge valuable discussions with A. Andreone, L. Cohen, N. Klein, S.Y. Lee, G. Müller, D.E. Oates, A. Pimenov, A. Purnell, and S. Sridhar.